\documentstyle[12pt,epsfig]{article}
\begin{document}
\begin{flushright}\hfill{\sf CERN-TH/2002-217}\end{flushright}
\begin{center}
{\Large Overproduction of primordial helium-4\\
in the presence of neutrino oscillations}\\
\ \\
D. P. Kirilova\footnote{Regular Associate of the Abdus Salam ICTP, 
Trieste, Italy}\\
\ \\
{\it Theory Division, CERN, Geneva, Switzerland and\\
    Institute of Astronomy, Bulgarian Academy of Sciences,\\
    blvd. Tsarigradsko Shosse 72, Sofia, Bulgaria\\
}
\end{center}

\begin{abstract}
The maximum overproduction of helium-4 in 
 cosmological nucleosynthesis
 with active--sterile neutrino oscillations,
$\nu_e \leftrightarrow \nu_s$,
efficient  after decoupling of electron neutrino, is analyzed.
The  kinetic effects on primordial nucleosynthesis due to 
neutrino spectrum distortion, caused by oscillations,
are precisely taken into account.\\
The maximum overproduction of primordial $^4\!$He as a function of 
oscillation parameters is obtained from the analysis of 
 the kinetics of the nucleons
and the oscillating neutrinos, for the full range of parameters of the
discussed oscillation model.
 A maximum relative increase  of  $^4\!$He,
up to $14\%$ for non-resonant oscillations and
up to $32\%$ for resonant ones  is  registered.
Cosmological  constraints on  oscillation parameters
are also discussed.
\end{abstract}
%
 
%

\newpage

\section{Introduction}

\indent  In this work I study the maximum overproduction of  $^4\!$He 
in   Big Bang
Nucleosynthesis (BBN) with  electron--sterile  neutrino oscillations
$\nu_e \leftrightarrow \nu_s$.

The positive indications for oscillations, obtained 
by the  neutrino
experiments (SuperKamiokande, SNO, Soudan 2, LSND, etc.)
  turned the subject of neutrino
oscillations
into one of the hottest points of astrophysics and neutrino physics.
The solar neutrino problem, the atmospheric neutrino anomaly and the
positive results of LSND experiments  may  be naturally resolved
by the phenomenon of neutrino oscillations, implying
nonzero neutrino mass and mixing. 
 Although  sterile neutrino impact in
oscillations explaining atmospheric and solar neutrino anomalies, is  
strongly constrained  from the analysis of experimental oscillations
data, still some small fraction of $\nu_s$ may participate in 
oscillations. Hence, it is interesting to study cosmological effects of 
such oscillations. 

Cosmological nucleosynthesis with neutrino oscillations
was studied in
numerous publications, discussed in
detail in ~\cite{dubnastro, dolgov}.
In these publications 
the central goal was to obtain cosmological
constraints on oscillation parameters. 

The overproduction of  $^4\!$He itself was not studied in detail 
until now.  
Such a study may be of interest for constructing of alternative 
cosmological models, for constraining galactic chemical evolution, 
it may be useful  for nonstandard models predicting active-sterile 
oscillations, like models with extra dimensions, mirror world 
particles, etc.   

Here I address that question of the possible maximal overproduction of 
$^4\!$He due to oscillations.   
For that purpose  the case of  $\nu_e \leftrightarrow 
\nu_s$,  oscillations,  effective after the   
electron neutrino decoupling from the plasma (i.e for
$\delta m^2 \sin^22\vartheta \le 10^{-7}$eV$^2$), is the most suitable. 
In that case due to the fact that 
sterile neutrino state is usually less populated than the active one 
at the start of oscillations, the oscillations may cause strong 
distortion of the electron neutrino spectrum, which affects the kinetics 
of the nucleons freezing before nucleosynthesis, and hence, the primordial 
production of  $^4\!$He. And, as will be shown by the numerical analysis, 
this kinetic effect of oscillations,  
may be much bigger than the  one corresponding to an increase in the 
total energy density due to  an additional neutrino flavor  $\delta 
N_{\rho}$, mainly 
considered in literature. 

We analyze $^4\!$He primordial production, taking into
account
 all known effects of $\nu_e \leftrightarrow \nu_s$
 oscillations on the primordial
synthesis of $^4\!$He.
 The production of  $^4\!$He  was calculated
 in the non-resonant and resonant oscillation cases.
In both cases strong overproduction of $^4\!$He was found possible
- up to $14\%$ and $32\%$, correspondingly. Thus in the discussed model of 
non-equilibrium oscillations the maximum overproduction of  $^4\!$He 
corresponds to an increase of the neutrino effective degrees 
of freedom  $\delta N_{kin}^{max}\sim 6$. 

The oscillation effects on BBN and  
their description are briefly reviewed 
 in the next section. The   kinetic approach,  
the results on  $^4\!$He primordially produced abundance  in the presence
of oscillations,  and the  cosmological constraints
on oscillation parameters are discussed in the last section.

\section{Cosmological Nucleosynthesis with Neutrino Oscillations}

\subsection{Standard Cosmological
Nucleosynthesis}

According to the standard BBN, during the early hot and dense
epoch of  the Universe,
the light elements
 D, $^3\!$He, $^4\!$He, $^7\!$Li were synthesized successfully.
The most
reliable and abundant data are now available  for $^4\!$He. 
This fact and  its relatively simple chemical evolution
make  $^4\!$He the preferred element
 for  the analysis of
the oscillations effect on BBN.

 The contemporary values for the mass fraction of $^4\!$He, $Y_p$,
 inferred from
observational data, are in the range  0.238--0.245
(the  systematic errors are supposed
to be around $0.007$)~\cite{izot}.

Primordially produced $^4\!$He abundance  $Y_p$,
is
calculated
with  great precision within the standard BBN model~\cite{lopez}.
According to the standard BBN  $^4\!$He  is
a result of a complex network of nuclear reactions,
proceeding after the  neutron-to-protons  freezing.
It  essentially
depends  on the freezing ratio $(n/p)_f$.  
The latter is a result of the freezing of the weak processes:
%

\begin{eqnarray}
\nu_e+n &\leftrightarrow& p+e^-
\nonumber\\
e^++n &\leftrightarrow& p+\tilde{\nu}_e,
\end{eqnarray}
which maintained the
equilibrium
of nucleons at high temperature ($T>1$ MeV).
Their freeze-out occurred when in the process of Universe cooling
the rates of these weak processes, $\Gamma_w$,
 became comparable and less than the expansion rate $H(t)$:

$$
 \Gamma_w \sim G_F^2 E_{\nu}^2 N_{\nu} \le H(t)\sim\sqrt{g_{eff}}~T^2
$$

 Thus, the produced $^4\!$He is a strong function of the
  number of the  effective degrees of freedom at BBN epoch,
$g_{eff}=10.75 +7/4\delta N_{\rho}$.
 

$Y_p$  depends also on the electron neutrino spectrum and 
number
density, and  on the neutrino-antineutrino 
asymmetry, which enter through $\Gamma_w$.
In the standard BBN model three neutrino
flavors, zero lepton asymmetry and equilibrium neutrino
number densities and spectrum distribution are postulated:

$$
 n_{\nu_e}(E)=(1+\exp(E/T))^{-1}.
$$

Almost all neutrons, present at the beginning of nuclear reactions,  are
sucked
into  $^4\!$He.
So, the primordially
produced mass fraction of  $^4\!$He can be
approximated by 

$$
 Y_p \sim2(n/p)_f/(1+n/p)_f \exp(-t/\tau_n).
$$
  
\noindent where   $\tau_{n}$ is the  neutron mean lifetime. 

The theoretical uncertainty of the precisely calculated $Y_p$
is less than  0.1\% ($|\delta Y_p| < 0.0002$) within a wide range of
$\eta$. So, in the standard BBN scenario, where $\tau_n$, and $g_{eff}$ are
fixed, $Y_p$, as well as the
rest cosmologically produced light elements 
 are functions of only one parameter - the baryon-to-photon ratio 
$\eta$.

Deuterium  measurements in  pristine
environments towards  low metallicity
quasar absorption systems  at very
high $z\sim3$  provide us with the most
precision determination of the baryon density
~\cite{maera},
giving the value: $\eta=(5.6 \pm 0.5) \times 10^{-10}$. Recently,
the baryon density was also determined 
 from observations of the anisotropy of the cosmic
microwave background (CMB) by DASI~\cite{pryke},
BOOMERANG~\cite{net,bernardis},
CMB~\cite{padin}
 and MAXIMA experiments~\cite{stompor}. For the combined analysis of 
the
data see also~\cite{wang,douspis}.
The CMB anisotropy data
is in remarkable agreement with the baryon density determined from
deuterium measurements and BBN.

So, the predicted primordial $^4\!$He abundance  $Y_p$ is in
accordance with the
observational data  and is consistent with deuterium
measurements.

\subsection{Effects of neutrino oscillations on nucleosynthesis}

In case neutrino oscillations are present in  the Universe primordial
plasma, they  lead to changes in the Big
Bang Nucleosynthesis, depending on the
 oscillation channels and the way they proceed.
The effect of flavor neutrino oscillations on BBN is
negligibly small because the temperatures and hence, the densities of the
neutrinos with different flavors are almost equal.

Active-sterile oscillations, however,  are capable to shift
 neutrino number densities and spectrum from their equilibrium values.
Besides, they may change
 neutrino-antineutrino asymmetry and
 excite additional neutrino types.
Thus,
the presence of
 neutrino oscillations invalidates  the main BBN assumptions
about
three neutrino flavors,
zero lepton asymmetry and
equilibrium neutrino  energy
distribution.

Qualitatively, the oscillations effects on the
nucleosynthesis of  $^4\!$He may be illustrated as follows:

\begin{itemize}
\item{excitation of additional degrees of freedom}
\end{itemize}
\indent It is known that active-sterile oscillations may keep
 sterile neutrinos in
thermal
equilibrium~\cite{dol} or bring them into equilibrium in case they have
already decoupled.
 The presence of light steriles in equilibrium leads to an increase
of the effective degrees of freedom during BBN and to
faster
Universe expansion  $H(t)\sim g_{eff}^{1/2}$  and
earlier $n/p$-freezing, $T_f\sim (g_{eff})^{1/6}$,
 at times when
neutrons were more abundant~\cite{dol,ns}:
$$
n/p\sim \exp\left[-(m_n-m_p)/T_f\right]
$$
 This effect leads to $5\%$ $^4\!$He overproduction corresponding 
to one additional neutrino type brought into equilibrium by 
oscillations. 

\begin{itemize}
\item{distortion of the neutrino spectrum}
\end{itemize}
\indent 
Much stronger effect of oscillations may be achieved in case of 
oscillations between initially empty sterile neutrino state and 
electron neutrino. The non-equilibrium initial condition leads to spectrum 
distortion of the active neutrino due to oscillations.

Since
oscillation rate is energy dependent $\Gamma \sim \delta m^2/E$
the low energy neutrinos start to oscillate first,
and later the
oscillations become noticeable for  the more energetic neutrinos. Due to
that,
the neutrino energy spectrum $n_{\nu}(E)$ may strongly deviate 
from its equilibrium form~\cite{kir}. This spectrum distortion 
  affects the kinetics of nucleons freezing - it leads to an earlier 
$n/p$-freezing and an overproduction  of
 $^4\!$He  yield. 

The effect  can be easily understood  having in mind that 
the  distortion leads both to a {\it depletion of the active neutrino
number densities} $N_{\nu}$~\cite{bd}:
$$
N_{\nu}\sim \int {\rm d}E E^2 n_{\nu}(E) 
$$

\noindent and to a {\it decrease of the mean neutrino energy}, 
which reflects into a
decrease of the weak rates  $\Gamma_w \sim E_{\nu}^2 N_{\nu}$,
and  hence, to  an overproduction of  $^4\!$He primordial
abundance.

The decrease of the electron neutrino energy due to
oscillations into low temperature sterile neutrinos,
has also an additional effect: Due to the threshold of
the reaction converting protons into neutrons,
when neutrinos have lower energy, protons are preferably produced 
in reactions (1), 
which may lead to an underproduction of  $^4\!$He~\cite{dk}.
However, this turns to be a minor effect.

The effect of spectrum distortion was
 numerically analyzed 
for hundreds combinations of mass differences and mixing angles in 
previous studies of 
 active-sterile oscillations ~\cite{kir,PL,res,PRD}.  
  It was proved
 important both in the resonant 
~\cite{res} and in the non-resonant oscillations case~\cite{PRD}.

In the discussed here  scenario it is the dominant effect and 
 leads to overproduction of the primordial
 $^4\!$He abundance.

\begin{itemize}
\item{neutrino-antineutrino
asymmetry growth}
\end{itemize}
\indent 
The idea of neutrino-antineutrino asymmetry generation
during the  resonant transfer of neutrinos was first  
proposed in ref.~\cite{smirnov}.
Dynamically produced asymmetry
 exerts back effect to
oscillating neutrino  and may change its
oscillation pattern~\cite{NU96,new}, it may suppresses
oscillations at small mixing angles, leading to
weakening  the  oscillations effect on BBN, i.e.
{\it  less overproduction of ~ $^4\!$He}.
\footnote{There were independent studies of the asymmetry growth in  
active-sterile 
oscillations for the case of large mass differences 
along the lines of the pioneer work~\cite{foot}.}

The effect of the oscillations generated asymmetry on
 $^4\!$He  for the discussed here model was
analyzed for hundreds of $\delta m^2-\vartheta$ combinations
in ~\cite{res}.

\section{Helium-4 overproduction due to
$\nu_e \leftrightarrow \nu_s$ neutrino oscillations}

We have provided an exact study of all the oscillation effects 
on the primordial production of  $^4\!$He. 

\subsection{The required  kinetic approach}

For the analysis of the
non-equilibrium picture of
active--sterile neutrino oscillations,  producing non-equilibrium
neutrino number densities, distorting   neutrino spectrum
and generating neutrino-antineutrino  asymmetry 
 a  self-consistent numerical analysis of the evolution of 
the
nucleons and the  oscillating neutrinos in the high temperature Universe
 was provided.

Exact kinetic equations for the nucleons and
for the neutrino density matrix {\it in momentum space}~\cite{PL}
  were used. This allowed to describe precisely
the kinetic effects of oscillations on helium production due to spectrum
distortion  {\it at each neutrino
momentum}. 

The equation for the  neutron number densities
in momentum space $n_n$ reads:
\begin{eqnarray}
&&\left(\partial n_n / \partial t \right)
 = H p_n~ \left(\partial n_n / \partial p_n \right) +
\nonumber\\
&& + \int {\rm d}\Omega(e^-,p,\nu) |{\cal A}(e^- p\to\nu n)|^2
\left[n_{e^-} n_p (1-\rho_{LL}) - \right.
\nonumber\\
&&\left. - n_n \rho_{LL} (1-n_{e^-})\right]
\nonumber\\
&& - \int {\rm d}\Omega(e^+,p,\tilde{\nu}) |{\cal A}(e^+n\to
p\tilde{\nu})|^2
\left[n_{e^+} n_n (1-\bar{\rho}_{LL}) - \right.
\nonumber\\
&& \left. - n_p \bar{\rho}_{LL} (1-n_{e^+})\right].
\end{eqnarray}
\noindent where ${\rm d}\Omega(i,j,k)$ is a phase space factor and
${\cal A}$ is the
amplitude of the corresponding process,  neutrino $\rho_{LL}$ and
antineutrino number densities $\bar{\rho}_{LL}$
at each integration step of eq.~(2) are taken  from the
simultaneously  performed integration of the set of equations
for neutrino density matrix (see ref.~\cite{PL}).

The equation  provides a simultaneous account of the different
competing processes,
namely: neutrino oscillations  (entering through  $\rho_{LL}$ and
$\bar{\rho}_{LL}$), Hubble expansion (first term) and weak  
interaction
processes (next terms).

The numerical analysis was  performed for the  temperature
interval  [$2$~MeV,  $0.3$~MeV].

We have found that for a wide range of oscillation parameters the 
spectrum distortion  of electron neutrino is considerable  
during the period of  nucleons freezing. Hence, usually the kinetic 
effect of oscillations due to electron neutrino energy spectrum 
distortion plays the dominant role in 
the overproduction of helium. 
In Fig.~1 the evolution of the energy spectrum of the electron  
neutrino through the period of nucleons freezing is illustrated. 
In the Figs.1a,1b and 1c the energy spectra at different 
characteristic temperatures, correspondingly $T=1$ MeV, 
 $T=0.7$ MeV and  $T=0.5$ are  presented.  By the dashed curve the 
equilibrium spectrum at the given temperature is given for comparison. 
The spectra are calculated for oscillation parameters 
$\delta m^2=10^{-7}$ eV$^2$ and $\sin^22\vartheta=0.1$. 

It is seen that oscillations affect non trivially the neutrino spectrum 
and distort strongly its equilibrium form. So, it is not possible to 
describe correctly the spectrum distortion due to oscillations 
simply by shifting the effective temperature of the neutrino 
and accounting only for the depletion of its total number density, 
considering its
spectrum equilibrium as in refs.~\cite{ssf,shif99}.  

For the proper description of the spectrum distortion in the 
case of non-equilibrium electron-sterile oscillations, studied here, 
the evolution of the neutrino ensembles should be explored using  
neutrino  density matrix in momentum space, allowing to 
describe the evolution of neutrino at each momentum.

The  neutrino-antineutrino asymmetry in the resonant case
 grows up to 5 orders of
magnitude from its initial value (taken to be of the order of
the baryon asymmetry). So, this asymmetry influences
BBN only indirectly - through oscillations.
This  oscillations generated asymmetry leads 
to a 
decrease in the produced $^4\!$He compared with the case without 
asymmetry account. However, its effect comprises only up to 
about a $10\%$ of the total effect. 

\subsection{Maximum helium-4 overproduction}

The overproduction of the primordial $^4\!$He,
$\delta Y_p=Y^{osc}_p-Y_p$ in the presence of
$\nu_e \leftrightarrow \nu_s$
 oscillations  was calculated for 
 mass differences
$\delta m^2 \le 10^{-7}$ eV$^2$ and 
$0\le\vartheta\le\pi/4$.
 Several hundreds of $\delta m^2-\vartheta$ combinations
were explored.

We have used the data from the 
 precise
calculations of the  $n/p$-freezing  provided  for the  non-resonant case 
in~\cite{PRD}
and for the
resonant case in~\cite{res}.
As far as it is the essential 
  for the
production of $^4\!$He.
The neutron decay was  accounted
adiabatically
 till the beginning  of
nuclear reactions at about $0.09$ MeV.

We have found that the effect of oscillations   becomes  very
small (less than $1\%$) for small mixings: as small as
$sin^22\vartheta=0.1$ for $\delta m^2=10^{-7}$ eV$^2$,
and  for small mass differences: $\delta m^2 <10^{-10}$ eV$^2$ at
maximal mixing.
For very small mass differences $\delta m^2 \le 10^{-11}$ eV$^2$,
or at very small mixing angles $\sin^22\vartheta \le 10^{-3}$,
the effect on nucleosynthesis becomes negligible.

In the non-resonant case the oscillation effect increases with the 
increase of the oscillation 
parameters, hence  
it  is maximal
at maximal mixing and  greatest mass differences.
In Fig.~2 (the lower curve) the
maximal  relative increase in the primordial $^4\!$He
as a function of neutrino mass differences at maximal mixing:
$\delta Y_p^{max}/Y_p=\delta Y^{osc}_p/Y_p(\delta
m^2)_{|\theta=\pi/4}$ is presented.

It is seen that for maximal mixing, the oscillation effect becomes 
greater than $5\%$ (the one corresponding to one additional neutrino 
type) already at $|\delta m^2|\ge 3\times 10^{-9}$ eV$^2$  (in the 
resonant 
case)  
and  $\delta m^2\ge 6 \times 10^{-9}$ eV$^2$ (in the non-resonant  
one). 
It continues to grow up with the increase of the mass differences, and 
at $|\delta m^2| \sim 10^{-7}$ eV$^2$ is several times bigger:  
$\delta N^{max}_{kin}\sim 3$ in the non-resonant case and $\delta 
N^{max}_{kin}\sim 6$ for the resonant one.

Further increase of the mass differences, however, will lead to 
oscillations effective before electron neutrino decoupling, 
and therefore, to a smaller spectrum distortion effect, because the 
interactions with the plasma will lead to faster thermalization of the 
sterile state. Hence, the effect on helium will decrease with 
further increase of  $|\delta m^2|$ and finally reach 
an overproduction of  
 $5\%$ again, corresponding to a full thermalization of the sterile state 
and its equilibrium spectrum.


In Fig.~3 we 
 present a combined plot (for the resonant and the non-resonant
oscillation case) of $\delta Y_p$
dependence on  the mixing angle for $\delta m^2=10^{-7}$ eV$^2$
 and  $\delta m^2=10^{-8}$ eV$^2$.
While in the non-resonant case the oscillations effect increases with the 
increase of the mixing (see l.h.s of Fig.3.),
in the resonant case
for a given $\delta m^2$
there exists some resonant mixing angle,
at which the oscillations
 are enhanced by the medium (due to the MSW effect), and hence,
the overproduction of
$^4\!$He is greater than that corresponding to the vacuum
maximal mixing angle. 
This behavior of the helium production on the
mixing angle is illustrated in the r.h.s. of the figure.

The upper curve in Fig.2  shows the
maximal relative increase  in the resonant
oscillations case  
 as a function of
 mass differences. Each maximum $^4\!$He value
  corresponds to the resonant mixing angle for the concrete mass
difference:  $\delta Y_p^{max}/Y_p=Y_p^{osc}(\delta 
m^2,\vartheta^{res}_{\delta m})$. As can be
seen from Figs.~2 and 3, a
considerable overproduction
can be achieved: in the resonant case up to $32\%$
  and in
the non-resonant one -- up to $14\%$.
So, the  net effect of spectrum distortion of 
electron neutrino due to oscillations on  
 the
production of  $^4\!$He  may
be considerable and 
several times larger than
 the effect 
due to excitation of
one  additional neutrino type.
 
Several words are due to 
cosmological  constraints on neutrino oscillation parameters,  
calculated in the discussed model of oscillations. 
Due to the strong kinetic effect of neutrino spectrum distortion 
caused by active-sterile oscillations discussed, the obtained constraints 
in that model of oscillations are more stringent than those obtained in 
pioneer works, not accounting for the kinetic  effects of 
oscillations. 
 Hence, according to these more precise studies of spectrum distortion 
effects of oscillations, the
$\delta Y_p/Y_p<3\%$ limit excludes almost completely the LOW
electron-sterile solution to the solar neutrino problem~\cite{now},
in addition to
the excluded sterile LMA solution in
previous investigations. I.e. it is more than an order of magnitude 
more restrictive to the mass differences.  
\ \\

This study has shown that considerable   
$Y_p$ overproduction may result from the 
electron neutrino spectrum distortion  due to
$\nu_e \leftrightarrow \nu_s$  oscillations. 
The overproduction is maximal for the case of initially empty sterile 
neutrino state, considered here. 
The dependence of  $^4\!$He overproduction 
on the degree of population of the sterile neutrino state before 
$\nu_e \leftrightarrow \nu_s$  oscillations is considered in a separate 
paper~\cite{spectrum}. Bigger initial population of $\nu_s$ leads to a 
smaller spectrum 
distortion of $\nu_e$, and hence smaller kinetic effect on primordial 
nucleosynthesis $\delta N_{kin}$ and higher energy density due to the 
increase of the number of degrees of freedom
 $\delta N_{\rho}$. The 
interplay between the two effects, however, is such that the 
overproduction of  $^4\!$He is  smaller than  in 
the case when initially  $\delta N_{\rho}=0$, discussed here.

\section{Conclusions}

 The primordial production of $^4\!$He in
the presence of $\nu_e \leftrightarrow \nu_s$ oscillations, effective 
after electron neutrino decoupling was analyzed.
 A precise quantitative study of 
the maximum overproduction of  $^4\!$He,  accounting for all oscillations
effects is provided. 
It was shown that 
the considerable spectrum distortion of the electron
neutrino caused by oscillations, which  effects the kinetics of
the neutron-proton transitions during nucleons freezing, 
plays the dominant role in the overproduction of  $^4\!$He.

 Enormous overproduction of $^4\!$He (up to
$32\%$ in the resonant case and $14\%$ in the non-resonant case)
was found possible  in case the sterile neutrino state was  empty at 
the 
start of  oscillations.

The results of this  analysis  can be useful  for constraining  
nonstandard 
physics,
 predicting active-sterile neutrino oscillations, like 
extra-dimensions, producing oscillations,
 supernova bursts employing oscillations, etc..
It  can be of interest also for models of galactic chemical evolution.

\section*{Acknowledgments}
\indent
I thank M. V. Chizhov for the help during   the preparation of
this paper.  
This work has been completed during my visiting position at TH CERN.
I appreciate  also the  Regular Associateship at the Abdus Salam ICTP, 
Trieste.

\newpage

\hspace{-1cm}\includegraphics[scale=0.8]{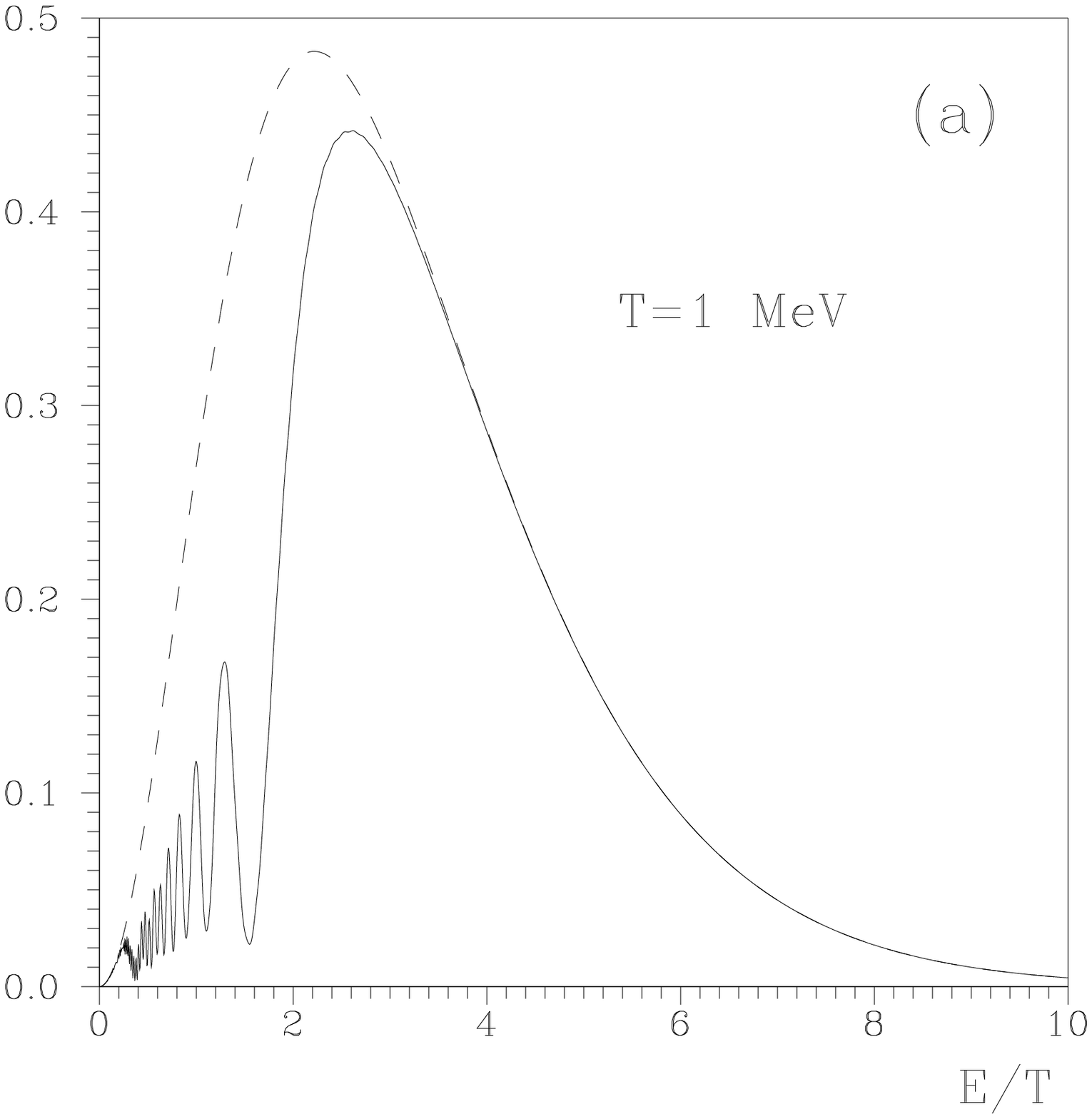}
\ \\

{\bf Figure 1a:}  The figure illustrates the degree of distortion
of the electron neutrino energy spectrum $x^2\rho_{LL}(x)$, where 
$x=E/T$,
caused by oscillations with mass difference
$|\delta m^2|=10^{-7}$ eV$^2$ and $\sin^22\vartheta=0.1$ at
 a characteristic temperature  $1$ MeV. The dashed curve gives the 
equilibrium neutrino spectrum for comparison.

\newpage

\hspace{-1cm}\includegraphics[scale=0.8]{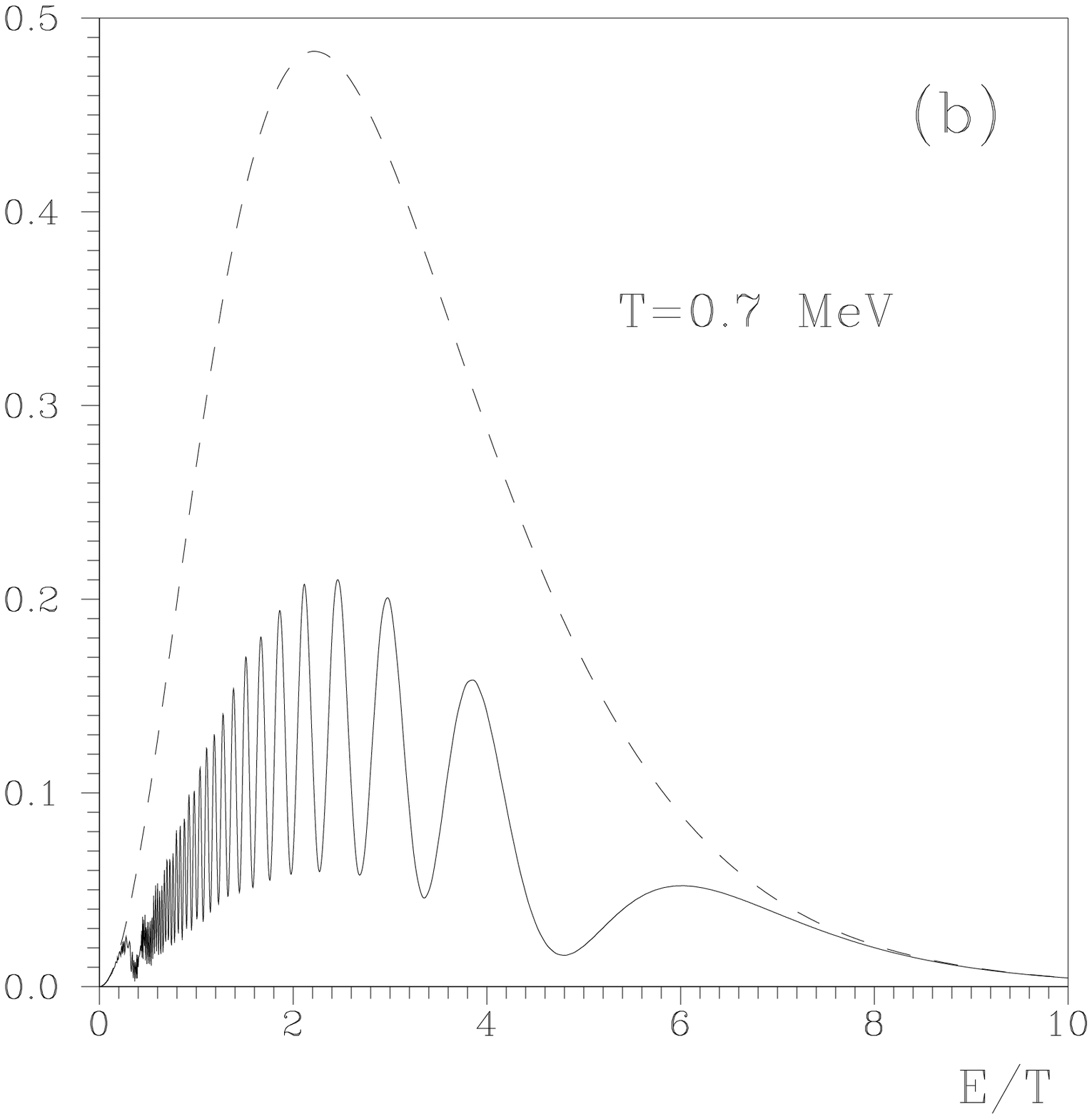}
\ \\

{\bf Figure 1b:} The figure illustrates the degree of distortion 
of the electron neutrino energy spectrum $x^2\rho_{LL}(x)$, where 
$x=E/T$, 
caused by oscillations with mass difference 
$|\delta m^2|=10^{-7}$ eV$^2$ and $\sin^22\vartheta=0.1$ at 
 a characteristic temperature $0.7$ MeV.

\newpage

\hspace{-1cm}\includegraphics[scale=0.8]{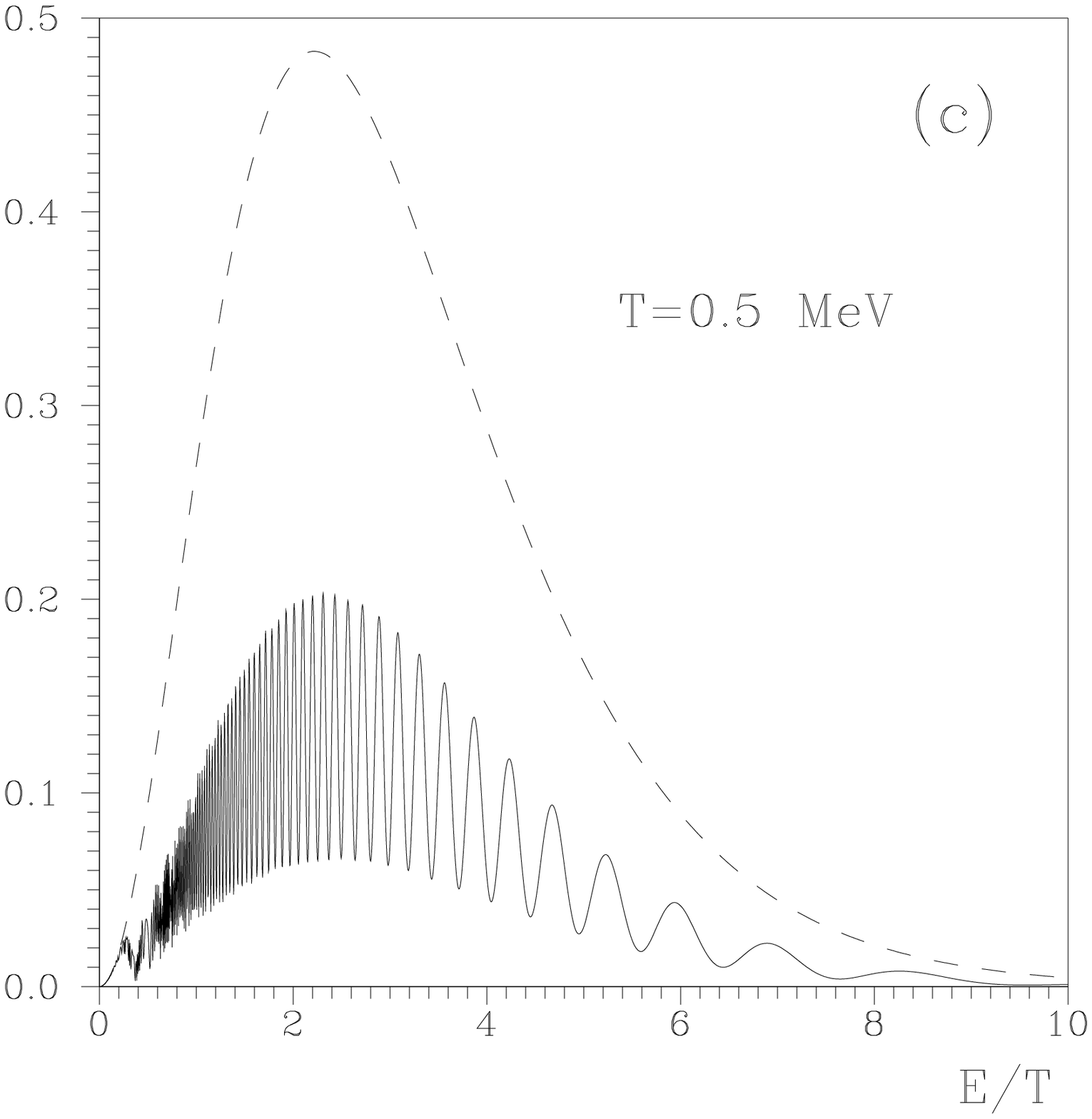}
\ \\

{\bf Figure 1c:} The figure illustrates the degree of spectrum distortion 
of the electron neutrino caused by oscillations with mass difference 
$\delta m^2=10^{-7}$ eV$^2$ and $\sin^22\vartheta=0.1$ at 
 a characteristic temperature $0.5$ MeV.

\newpage

\hspace{-1cm}\includegraphics[scale=0.8]{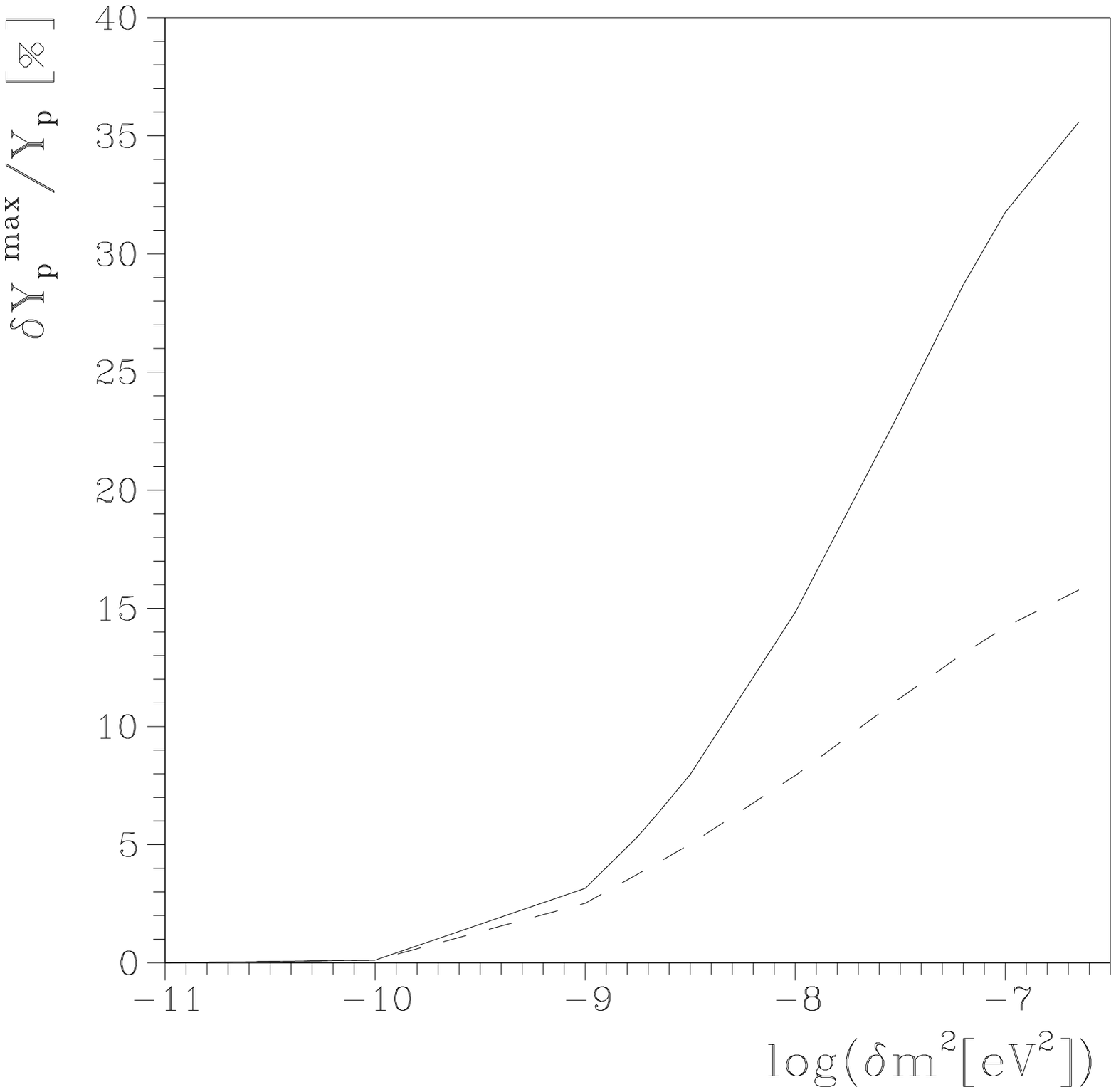}
\ \\

{\bf Figure 2:} Maximum primordial  $^4\!$He abundance
for the resonant  (upper curve)
and the non-resonant oscillation case (lower curve), as a function
of the neutrino mass differences. The non-resonant case is calculated at
maximum mixing, while in the resonant case the
 helium abundance is calculated at the resonant mixing angle
for
the corresponding  mass difference.

\newpage

\hspace{-1cm}\includegraphics[scale=0.8]{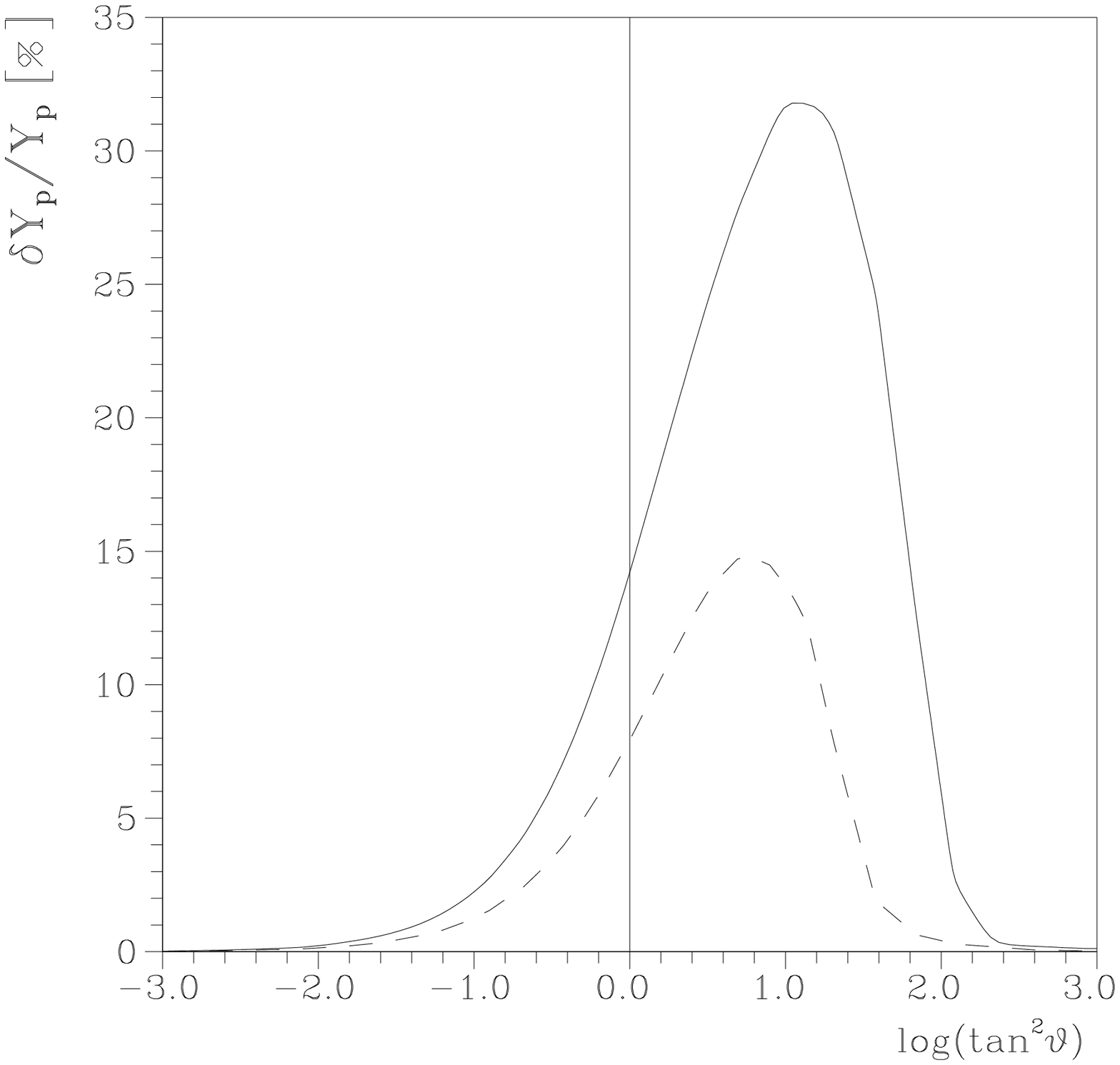}
\ \\

{\bf Figure 3:} The dependence of the relative increase
of  primordial helium  on the mixing angle
for the resonant (r.h.s.) and non-resonant (l.h.s.) oscillation
case.
The upper curve corresponds to $\delta m^2=10^{-7}$ eV$^2$,
the lower one to $\delta m^2=10^{-8}$ eV$^2$.

\newpage

\end{document}